\documentclass[%
 aip,
 amsmath,amssymb,
 reprint,%
]{revtex4-1}
\usepackage{dcolumn}
\usepackage{bm}
\usepackage[utf8]{inputenc}
\usepackage[T1]{fontenc}
\usepackage{mathptmx} 
\usepackage[normalem]{ulem}
\usepackage{mathtools}
\usepackage{graphicx}
\usepackage{epstopdf}
\usepackage{lmodern}
\usepackage{xfrac}
\usepackage[breaklinks]{hyperref}
\usepackage{mathtools}
\usepackage{xcolor}

\begin{document}

\title{Absolute clock synchronization with a single time-correlated photon pair source over 10\,km}

\author{Jianwei~Lee}
\affiliation{Centre for Quantum Technologies, National University of
  Singapore, 3 Science Drive 2, Singapore 117543, Singapore}

\author{Lijiong~Shen}
\affiliation{Centre for Quantum Technologies, National University of
  Singapore, 3 Science Drive 2, Singapore 117543, Singapore}
\affiliation{Department of Physics, National University of Singapore, 2
  Science Drive 3, Singapore 117551, Singapore}

\author{Adrian Nugraha Utama}
\affiliation{Centre for Quantum Technologies, National University of
  Singapore, 3 Science Drive 2, Singapore 117543, Singapore}

\author{Christian~Kurtsiefer}
\email{christian.kurtsiefer@gmail.com}
\affiliation{Centre for Quantum Technologies, National University of
  Singapore, 3 Science Drive 2, Singapore 117543, Singapore}
\affiliation{Department of Physics, National University of Singapore, 2
  Science Drive 3, Singapore 117551, Singapore}
\date{\today}
\begin{abstract} 
We demonstrate a point-to-point clock synchronization protocol based on
bidirectionally propagating photons generated in a single spontaneous parametric
down-conversion (SPDC) source.
Tight timing correlations between photon pairs are used to determine the
single and round-trip times measured by two separate clocks, providing
sufficient information for distance-independent absolute synchronization secure against symmetric delay attacks. We show that the coincidence signature useful for determining the round-trip time of a synchronization channel, established using a 10\,km telecommunications fiber, can be derived from photons reflected off the end face of the fiber without additional optics. Our technique allows the synchronization of multiple clocks with a single reference clock co-located with the source, without requiring additional pair sources, in a client-server configuration suitable for synchronizing a network of clocks.
\end{abstract}
\maketitle


\emph{Introduction - }Complementary to clock recovery schemes from data streams, absolute clock synchronization protocols, e.g. network time protocol (NTP), precision time protocol (PTP), two-way satellite time
transfer (TWSTT), are widely-used to determine the offset between physically separated
clocks~\cite{wenjun2014two,mills:1991,PTP,Moreira:2009}. 
By exchanging counter-propagating signals, and assuming a symmetric synchronization channel, parties estimate one-way propagation delays as half the round-trip time signals without characterizing their physical separation beforehand. 
Spatially separated parties then deduce their absolute clock offset by comparing signal propagation times measured with their devices with the expected propagation delay~\cite{narula:17}. 
Recently, protocol implementations with entangled photon pairs suggest securing the synchronization channel by measuring non-local correlations -- a technique inspired by entanglement-based quantum key distribution (QKD)~\cite{lee2019symmetrical,hou2018fiber,lamas2018secure}. 
However, to realize a bidirectional exchange of photons, these demonstrations required a photon pair source at each end of the synchronization channel, posing a resource challenge when synchronizing multiple clocks.

In this work, we experimentally demonstrate a bidirectional clock synchronization protocol where the synchronization channel is established with a 10\,km optical fiber and a single entangled photon pair source. 
The round-trip time is sampled using time-correlation measurements between the detection times of photon pairs, with one photon of the pair back-reflected at the remote side using the end face of the fiber. 
We demonstrate a distance-independent synchronization of two separated clocks, referenced to independent rubidium frequency standards. 
Already from a quite modest photon pair detection rate of $160\,$s$^{-1}$
we obtain a precision sufficient to resolve clock offset fluctuations with an uncertainty of 88\,ps in 100\,s, consistent with the intrinsic frequency instability between our clocks.

\begin{figure}
  \centering
  \includegraphics[width=\columnwidth]{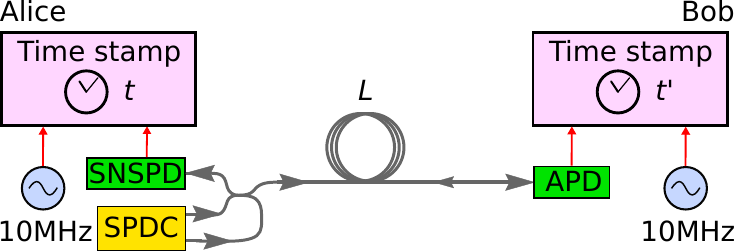}
  \caption{\label{fig:setup}
    Clock synchronization setup. Alice has a source of time-correlated photon pairs based on spontaneous parametric down-conversion (SPDC) and a single-photon nanowire photodetector (SNSPD). 
    One photon of the pair is detected locally, 
    while the other one is sent through a single mode fiber of length~$L$ to be detected on the remote side with Bob's InGaAs avalanche photodiode (APD).
    Times of arrival for all detected photons are recorded at each side with respect to the local clock, each locked to a rubidium frequency reference (10\,MHz).
    Occasionally, a transmitted photon is reflected at the end face of the fiber back to Alice, allowing her to determine the round-trip time and derive the absolute offset between the clocks.
  }
\end{figure}
   
\begin{figure}
\includegraphics[width=\columnwidth]{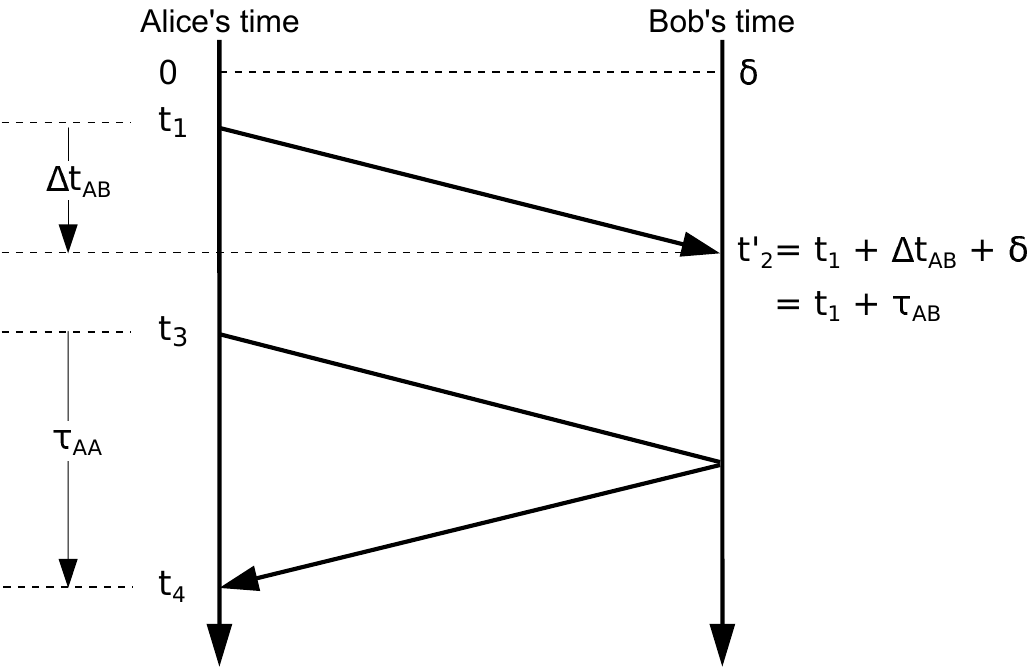}
\caption{\label{fig:syncscheme}
  Clock synchronization scheme. 
  Alice and Bob measure detection times $t$ and $t'$ of photon pairs generated from Alice's source using local clocks.
  Detection times $t_1$ and $t'_2$ are associated with a time-correlated photon pair where one photon of the pair is transmitted to Bob, while $t_3$ and $t_4$ are associated with a pair where one of the photons is reflected at Bob back to Alice.
  The single-trip time $\tau_{AB}$ of photons in the synchronization channel, calculated from the time difference $t'_2-t_1$, depends on the signal delay $\Delta t_{AB}$ associated with the length of the channel, and the absolute clock offset $\delta$ between the clocks. 
  The round-trip time $\tau_{AA}$ of the channel is estimated using $t_4-t_3$.
  Assuming a symmetric delay channel, $\delta$ can be derived from $\tau_{AB}$ and $\tau_{AA}$ without \textit{a priori} knowing $\Delta t_{AB}$. 
}   
\end{figure}

\emph{Time synchronization protocol -}
The protocol involves two parties, Alice and Bob, connected by a single mode optical fiber~(see Fig.~\ref{fig:setup}). 
Alice has an SPDC source producing photon pairs, one photon is detected locally, while the other is sent and detected on the remote side.
Occasionally, the transmitted photon undergoes Fresnel reflection ($R\approx3.5\%$) at the end face of the fiber, and is eventually detected by Alice instead.
Every photodetection event is time tagged according to a local clock which assigns time stamps $t$ and $t'$ at Alice and Bob, respectively. 

Photon pairs emerging from SPDC are tightly time-correlated~\cite{hong1987measurement}.
Thus, for an offset $\delta$ between the clocks,  
a propagation time $\Delta t_{AB}$ from Alice to Bob, 
and $\Delta t_{BA}$ in the other direction, 
the second-order correlation function~\cite{glauber1963quantum} $G^{(2)}(\tau)$ of the time difference $\tau = t'-t$ has a peak at
\begin{equation}
  \tau_{AB} = \delta + \Delta t_{AB}
\end{equation}
due to pairs detected at opposite ends of the channel, 
whereas for two photons detected by Alice at $t$ and $t+\tau$, the auto-correlation function $R(\tau)$ will show a peak at
\begin{equation}\label{eq:round-trip time }
  \tau_{AA} = \Delta t_{AB} + \Delta t_{BA},
\end{equation}
corresponding to the round-trip time  of the channel.
If the propagation times in the two directions are the same, $\Delta t_{AB}=\Delta t_{BA}$, 
the the clock offset
can be deduced directly from the positions of the two peaks using
\begin{equation}\label{eq:offset}
\delta = \tau_{AB}  - \frac{1}{2}\,\tau_{AA},
\end{equation}
independently of the propagation time $\Delta t_{AB}$.
In this way, the protocol is inherently robust against symmetric changes in channel propagation times.

\emph{Experiment -}
A sketch of the experimental setup is shown in Fig.~\ref{fig:setup}. 
Our photon pair source~\cite{shi2020stable,lohrmann2020broadband,grieve2019characterizing} is based on Type-0 SPDC in a periodically-poled crystal of
potassium titanyl phosphate (PPKTP) pumped by a laser diode
at 658\,nm (Ondax, stabilized with holographic grating). 
The resulting photon pairs are degenerate at 1316\,nm, close to the zero dispersion wavelength of the  synchronization channel (SMF-28e, $10$\,km), with a bandwidth of $\approx50\,$nm on either side of this wavelength~\cite{grieve2019characterizing}.
Signal and idler photons are efficiently separated using a wavelength division
demultiplexer (WDM). 
Fiber beam splitters separate the photon pairs so that one photon is detected locally with a superconducting nanowire single-photon detector (SNSPD, optimized for 1550\,nm), while the other photon is routed into the synchronization channel where it is detected on the remote side with an InGaAs avalanche photodiode (APD).
The SNSPD has relatively low jitter ($\approx 40\,$ps) compared to APDs, and allows Alice to measure the round-trip time more accurately regardless of the choice of detector by the remote party.
With a pump power of 2.5\,mW focused to a beam waist of $140\,\mu$m at the centre of the crystal, we observed pair rates of $160\,$s$^{-1}$ and $8900\,$s$^{-1}$ associated with the round-trip and single-trip propagation of photons, respectively.

\begin{figure}
  \centering
  \includegraphics[width=\columnwidth]{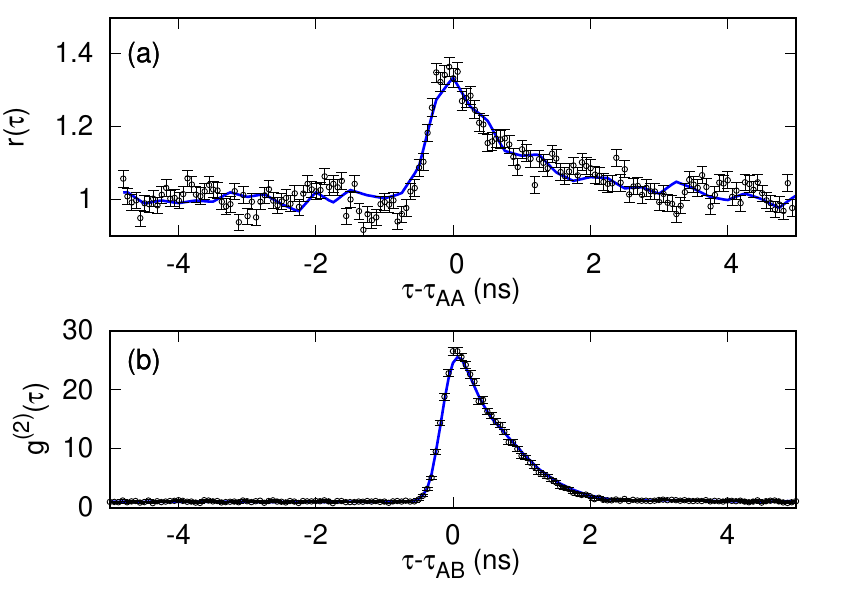}
  \caption{\label{fig:fits}
  Timing correlations showing coincidence peaks due to (a) round-trip and (b) single-trip propagation of photons in the synchronization channel.
  (a) $r(\tau)$: auto-correlation function $R(\tau)$ normalized to background coincidences extracted from Alice's timestamps acquired over 90\,s.  
  (b)  $g^{(2)}(\tau)$: cross-correlation function $G^{(2)}(\tau)$ normalized to background coincidences extracted from Alice and Bob's timestamps acquired over 3\,s.
  Solid lines: fits to heuristic model.
  $\tau_{AA}$ and $\tau_{AB}$: peak positions of respective distributions.
  Error bars: propagated Poissonian counting statistics.
  }
\end{figure}

Photon detection times $t$ and $t'$ at Alice and Bob are registered with a nominal resolution of $\approx$4\,ps.
We compute~\cite{ho2009clock} the histograms $G^{(2)}(\tau)$ and $R(\tau)$ with a bin width of of 62.5\,ps, and observed coincidence peaks associated with the single-trip and round-trip propagating photons (FWHM = 905\,ps and 950\,ps, respectively).
Figure~\ref{fig:fits} shows the respective histograms normalized to background
coincidences when the two clocks are locked to a common rubidium frequency
reference (Stanford Research Systems FS725), 
seperated by a fiber spool of constant length $L = 10$\,km. 
To deduce the clock offset, we first generate empirical models (Fig.~\ref{fig:fits}, solid-lines) for the two coincidence peaks using 100\,s of timestamp data --
the models are used to fit subsequent histograms to extract peak positions $\tau_{AB}$ and $\tau_{AA}$.
With the peak positions, we then determine the clock offset using Eqs.~\ref{eq:round-trip time } and~\ref{eq:offset}.

\begin{figure}
  \centering
  \includegraphics[width=\columnwidth]{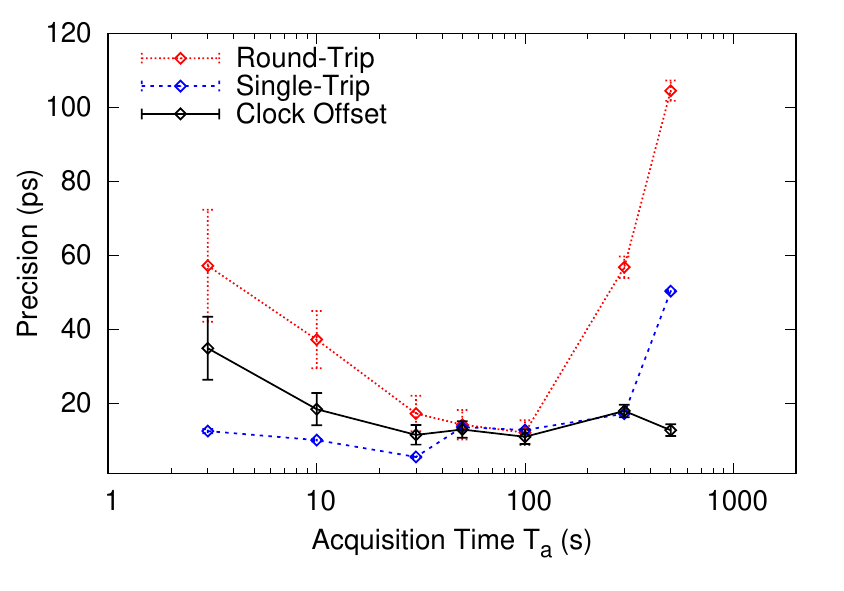}
  \caption{\label{fig:precision}
  Precision of the round-trip (red) and single-trip (blue) times, and the clock offset (black) between two clocks. 
  Both clocks are locked to the same frequency reference.  
  Error bars: precision uncertainty due to errors in determining the positions, $\tau_{AB}$ and $\tau_{AA}$, of the coincidence peaks.
  }
\end{figure}
To characterize the synchronization precision $\delta t$ as a function of the acquisition time, we measure the standard deviation of twenty offset measurements, each extracted from time stamps recorded for a duration $T_a$.
Figure~\ref{fig:precision} shows the precision of the measured offset, single-trip ($\tau_{AB}$) and round-trip times ($\tau_{AA}$).
We observe that the precision for the single and round-trip times improves with $T_a$ for timescales $\lesssim100$\,s, 
but deteriorates for longer timescales. 
We attribute this effect to temperature-dependent ($\Delta T=45\,$mK over 1\,min, $160\,$mK over 3\,hours) length fluctuations, given that the propagation delay variation~\cite{bousonville2009velocity} of our fiber is several 10\,ps\,km$^{-1}$\,K$^{-1}$.
However, we observe that these long-term fluctuations are suppressed in the clock offset measurement with the distance-independent synchronization protocol.

For subsequent demonstrations, 
we set $T_a=3$\,s and $90\,$s for the single and round-trip time measurements, obtaining a precision of 
12\,ps and 14\,ps, respectively.
Each 90\,s window used to evaluate the round-trip time thus contains thirty single-trip time measurements.
For each single-trip time value, we evaluate the clock offset using the round-trip time evaluated in the same window.
This results in a precision of 16\,ps for the measured offset.
Measuring the single-trip delay with shorter $T_a$ enables frequent measuring of $G^{(2)}(\tau)$, and is useful for tracking the position of its coincidence peak ($\tau_{AB}$) in the scenario where  clocks are locked to independent frequency references. 

The minimum resolvable clock separation associated with the offset precision is $3.3$\,mm.
To demonstrate that the protocol is secure against symmetric channel delay
attacks, we change the propagation length over several meters during
synchronization --- three orders of magnitude larger than the minimum resolvable length-scale.  

\emph{Distance-independent clock synchronization with the same reference clock
  - }
\begin{figure}
  \centering
  \includegraphics[width=\columnwidth]{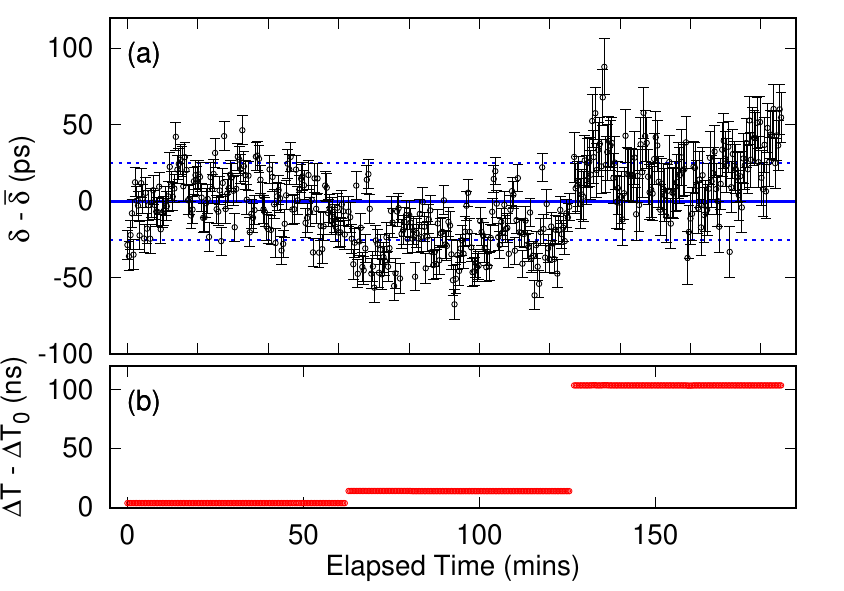}
  \caption{\label{fig:offsetvsdist}
  (a) Measured offset $\delta$ between two clocks, both locked on the same frequency reference. 
  The continuous line indicates the average offset $\overline{\delta}$.
  Error bars: precision uncertainty due to errors in determining the positions, $\tau_{AB}$ and $\tau_{AA}$, of the coincidence peaks.
  Dashed lines: one standard deviation. 
  (b) The round-trip time $\Delta T$ was changed using fiber lengths $L=L_0=10\,$km, $L_0+1\,$m, and $L_0+10\,$m. $\Delta T_0 = 103.3\,\mu$s.
  }
\end{figure}
To simulate a symmetric channel delay attack, we
impose different propagation distances using different fiber lengths. 
Figure~\ref{fig:offsetvsdist} shows the measured offset $\delta$ and the round-trip time $\Delta T$, with an overall standard deviation of 
26\,ps,
and an overall mean of $\bar{\delta}$.
The sets of $\delta$ obtained for $L = L_0+1\,$m and $L_0+10\,$m, 
with mean offsets 
$\bar{\delta}-24(17)\,$ps, and $\bar{\delta}+20(20)\,$ps, 
respectively, show significant overlap with those obtained with 
$L=L_0=10$\,km with mean offset 
$\bar{\delta}+1(17)\,$ps.
Comparing the additional mean offset of 19(26)\,ps to the additional single-trip delay ($48.3\,$ns) expected for extending our optical channel from $L=L_0$ to $L_0+10\,$m, our protocol suppresses the contribution of the additional propagation delay on the measured offset by a factor of $\approx 4\times10^{-4}$.
\emph{Distance-independent clock synchronization with independent clocks -}
\begin{figure}
  \centering
  \includegraphics[width=\columnwidth]{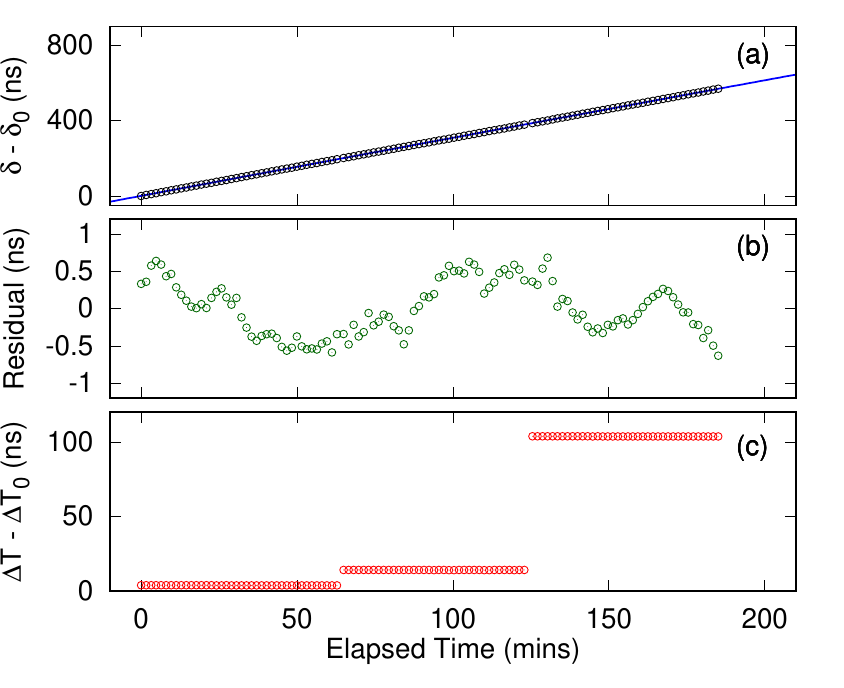}
  \caption{\label{fig:offsetvsdistsepclocks}
  (a) Measured offset $\delta$ between two clocks with different frequency references. 
  Each value of $\delta$ was evaluated from measuring photon pair timing correlations for 3\,s.
  The offset measured at the beginning is $\delta_0$. 
  Continuous blue line: fit used to extract the relative frequency accuracy ($\approx5.16\times10^{-11}$) between the clocks.
  (b) Residual of the fit fluctuates due to the intrinsic instability of the individual frequency references. 
  (c) The round-trip time $\Delta T$ was changed using three different fiber lengths.
  }
\end{figure}
To examine a more realistic scenario, we provide each time-stamping unit with an independent frequency reference (both Stanford Research Systems FS725), 
resulting in a clock offset that drifts with time~$\delta\rightarrow\delta(t)$. 

The frequency references each have a nominal frequency accuracy $d_0 < 5\times10^{-11}$, resulting in a relative accuracy $\sqrt{2}\,d_0$ between two clocks.
We evaluate the offset from the time stamps every $T_a = 3$\,s so that the maximum expected drift ($<$212\,ps) of the coincidence peak in $G^{(2)}(\tau)$ is smaller than its FWHM.
This pseudo-stationary regime allows the peak positions to be extracted with the same fitting procedure used when the clocks are locked onto the same frequency reference~\cite{lee2019symmetrical}.

We again simulate a symmetric channel delay attack using three different values of $L$.
Figure~\ref{fig:offsetvsdistsepclocks} shows the measured $\delta(t)$ which appears to follow a continuous trend over different round-trip times, indicating that the delay attacks were ineffective. 
Discontinuities in $\delta(t)$ correspond to periods when fibers were changed. 

To verify that meaningful clock parameters can be extracted from $\delta(t)$ despite the attack, we fit the data to a parabola $a\,t^2 + d\,t + b$, where $a$, $d$ and $b$ represent the relative aging, frequency accuracy and bias of the frequency references, respectively~\cite{Xu:2016ji}. 
The resulting relative frequency accuracy between the clocks, $d = 5.1654(7)\times10^{-11}$, agrees with the nominal relative frequency accuracy $\sqrt{2}\,d_0$ between our frequency references.
The residual of the fit, $r(t)$ (Fig.~\ref{fig:offsetvsdistsepclocks}(b)), 
fluctuates~\cite{riley2008handbook}~(Allan deviation~$=2.2\times10^{-12}$, time deviation~$=88\,$ps  in 100\,s) mainly due to the intrinsic instabilities of our frequency references~($2\times10^{-12}$ in 100\,s each). 

\emph{Protocol Security -}
Although not demonstrated in this work, Alice and Bob can verify the origin of each photon by synchronizing with polarization-entangled photon pairs and performing a Bell measurement to check for correspondence between the local and transmitted photons. 
This proposal addresses the issue of spoofing in current classical synchronization protocols~\cite{lee2019symmetrical,lamas2018secure}. 
However, due to low coincidence-to-accidental ratio  associated with the round-trip time  measurement (CAR=0.13), this authentication scheme is only feasible for the single-trip time measurement (CAR=8.9). 
Consequently, users can only authenticate photons traveling from Alice to Bob, and have to assume that the synchronization channel has not been asymmetrically manipulated in order to
incorporate the round-trip time measurement in the clock offset calculation (Eqn.~\ref{eq:offset}).

In addition, we also assumed that the photon propagation times in both directions were equal ($\Delta t_{AB} = \Delta t_{BA}$).
Without this assumption, the offset
\begin{equation}
  \delta = \tau_{AB} - \tau_{AA} + \Delta t_{BA}
\end{equation} 
can no longer be obtained directly from the peak positions $\tau_{AB}$ and $\tau_{AA}$. 

We note that an adversary will be able to exploit both assumptions while evading detection by passively rerouting photons traveling in opposite directions in the synchronization channel without disturbing their polarization states~\cite{lee2019asymmetric}. 
This attack is based on the fact that the momentum and polarization degree-of-freedoms of our photons are separable, and remains a security loophole in similar implementations~\cite{lee2019symmetrical,hou2018fiber}.

\emph{Conclusion -}
We have demonstrated a protocol for synchronizing two spatially separated clocks absolutely 
with time-correlated photon pairs generated from SPDC.
By assuming symmetry in the synchronization channel,
the protocol does not require \textit{a priori} knowledge of the relative distance or propagation times between two parties, providing security against symmetric channel delay attacks and timing signal authentication via the measurement of a Bell inequality~\cite{lamas2018secure}.
Compared to previous implementations~\cite{lee2019symmetrical,hou2018fiber},
our protocol requires only a single photon pair source, relying on the
back-reflected photon to sample the round-trip time of the synchronization
channel. This arrangement allows multiple parties to synchronize with
bidirectional signals with a single source.

With our protocol, we synchronize two independent rubidium clocks while changing their relative separation, using telecommunication fibers of various lengths ($\geq10\,$km) as a synchronization channel. 
Even with relatively modest detected coincidence rates ($160\,$s$^{-1}$) used for the round-trip time measurement, 
we obtained a precision sufficient to resolve clock offset fluctuations with a time deviation of 88\,ps in 100\,s, consistent with the intrinsic frequency instabilities of our clocks.
The precision improves with detectors with lower timing
jitter~\cite{hou2018fiber}, brighter sources, or for a transmission channel   
with insignificant dispersion (free space).
Frequency entanglement may also be leveraged to cancel dispersion non-locally, improving protocol precision over optical channels in future work~\cite{hou2018fiber}.

We thank S-Fifteen Instruments for assistance with the entangled photon pair source and the InGaAs detector.
This research is supported by the National Research Foundation, Prime Minister’s 
Office, Singapore and the Ministry of Education, Singapore under the Research 
Centres of Excellence programme.

\bibliographystyle{apsrev4-1}
%

\end{document}